\documentclass[twocolumn,prb,color,psfig,showkeys,superscriptaddress]{revtex4}
\usepackage{graphicx}
\usepackage{amsmath}
\usepackage{amssymb}
\newcommand{\eps}{\varepsilon}
\newcommand{\f}{\varphi}
\newcommand{\w}{\omega}

\newcommand{\ket}[1]{\ensuremath{| #1 \rangle}}
\newcommand{\bracket}[2]{\ensuremath{\langle #1 | #2 \rangle}}
\newcommand{\elem}[3]{\ensuremath{\langle #1 | #2 | #3 \rangle}}

\begin{document}

\title{The model of the optical switching center \\ and the polarization anomalies in
absorption spectra of cupric oxide after fast particles irradiation}

\author{E.V. Zenkov}
\email{eugene.zenkov@mail.ru}
\affiliation{Ural State Technical University, 620002 Ekaterinburg, Russia}

\begin{abstract}
We consider the model of the optical switching center -- a system with the following
properties: It has two (or more) metastable states $\ket{1}$, $\ket{2}$, separated by
a potential barrier $U$; it can switch from one state to another by absorbing the
photons with energy $\hbar \w \sim U$; the transition $\ket{1} \rightarrow \ket{2}$ is
allowed only for a certain light wave polarization $p_1$ and the transition $\ket{2}
\rightarrow \ket{1}$ -- for other polarization $p_2$; these polarizations $p_1$, $p_2$
are orthogonal. The optical properties of this system are studied and are found to
exhibit unconventional polarization dependence. In particular, the absorption spectrum
observed in natural (unpolarized) light can display new features, that are absent in
the spectra, obtained in two independent polarizations. We discuss these results in
connexion with the (yet unexplained) experimental findings \cite{CuHe}, where the
similar anomalous polarization dependence of the absorption spectra of cupric monoxide
CuO after the fast particle bombardment is reported.
\end{abstract}


\maketitle

\section{Introduction}

The copper oxides and related materials show a number of unconventional physical
properties, that have placed them among the most studied systems in the past decade.
Of especial interest is their liability to the formation of various metastable and
nonuniform states. Examples include the phase separation phenomena in doped HTSC
cuprates \cite{uemura}, the stripe structures in HTSC's and non-superconducting
oxides, such as the cupric monoxide CuO \cite{prl85}, the catalytic properties of CuO
in certain chemical processes \cite{Ohyama}. A general conclusion to be drawn from
these facts is that the equilibrium state of the systems may be readily disturbed by
means of various perturbations, such as chemical doping.

An important information about the concomitant changes of the electronic structure of
these systems can be obtained in optical studies. In a series of papers
\cite{CuHe,CuEl,CuAzot} the evolution of the optical absorption spectra of CuO single
crystals under the bombardment with the fast particles have been studied in a
systematic way. It was found, that the bombardment leads to the progressive growth of
the spectral weight within the dielectric gap and the appearance of new peaks in
absorption spectra, that have no counterparts in the spectra of the non-irradiated
good single crystalline samples. Similar results have been reported recently for the
absorption spectra of CuO powder with nanosize grains after the influence of the
converging spherical shock waves \cite{gizhevskii}.

Overall, these findings can be interpreted within the picture of the phase separation,
stimulated with the fast particles bombardment or the shock waves. The new spectral
features, emerging in these conditions can possibly be attributed to the surface
plasmon (Mie) resonances, as it was shown in \cite{gizhevskii}. Although the
effective medium model of \cite{gizhevskii} argue the novel phase to be presumably
''metallic'', the microscopic physics of cuprates in the phase separation regime
remains an issue and any additional detailed experimental data concerning the optical
properties of these systems is of especial interest.

The present work is motivated with the investigations of the absorption spectra of CuO
single crystal, irradiated with the fast He$^+$ nuclei, in different polarizations
\cite{CuHe}. Surprisingly, it was found, that the spectrum in unpolarized light looks
quite differently from the spectra in two plane polarizations and shows a new feature,
not observed in any of these polarizations. To the best of our knowledge, no similar
results have been ever reported for other systems. We interpret these findings as a
manifestation of a subtle kind of optical nonlinearity, which origin is related to the
effect of the bobmardment and the concomitant nucleation of some novel phase. In
general, the nonlinearity take place each time as the system modifies its properties
(susceptibilities, etc.) under the influence of an external perturbation, so that the
response is no longer proportional to the magnitude of the perturbation. Thus, it is
plausible, that the metastability and phase separation may favour the enhancement
nonlinear phenomena.

The article is organized as follows. The next section contains a short overview of the
experiment and its discussion. In sections 3, 4 we formulate our model and examine its
optical properties in different polarizations. In section 5 the microscopic grounds of
the model and its application to cuprates are considered. In section 6 the summary of
the work is given.

\section{Experiment}

As the experimental findings under consideration have been published elsewhere
\cite{CuHe}, we restrict ourself here to the most essential facts. The CuO single
crystalline platelet, cut out in crystallographic $ac$ plane, have been exposed to the
bombardment with the He$^+$ nuclei and the absorption coefficient of the irradiated
sample have been measured within the range of 0.2 -- 1.4 eV, the light propagating
normally to the platelet. The bombardment was found to result in a significant
destruction of the fundamental band edge and the appearance of some novel sharp peaks
within the range 0.9 -- 1.1 eV, that have no counterpart in the spectra of good CuO
single crystals, shown at the inset of figure 1.

Main results are summarized at the figure 1. It is seen, that the change of the
polarization of incident light from $\mathbf{E} \parallel [\overline{1}01]$ to
$\mathbf{E} \perp [\overline{1}01]$ leads to the shift of the peak from $\sim$ 0.9 eV
to $\sim$ 1.1 eV. Taking into account the anisotropy of the cupric monoxide, this
rather strong dependence of the spectra on the polarization is not surprising.
However, really striking is the spectrum, obtained in the unpolarized light, that
shows again the similar peak shifted to still higher frequency ($\sim$ 1.3 eV).

The origin of these new spectral features and especially their relative positions in
polarized and natural light pose a challenging problem. It is clear, that as far as
the medium is linear, the spectrum, obtained in natural light, should always reduce to
the superposition of the spectra, obtained in any two different polarizations,
whatsoever are the mechanisms of the light extinction within the medium. Indeed, the
unpolarized light may be regarded as a random mixture of two light waves with
different polarizations. If the waves travel through the sample independently, the
resulting optical spectrum will display the features, arising from both waves.
\begin{figure}[t]
\includegraphics[width=\linewidth]{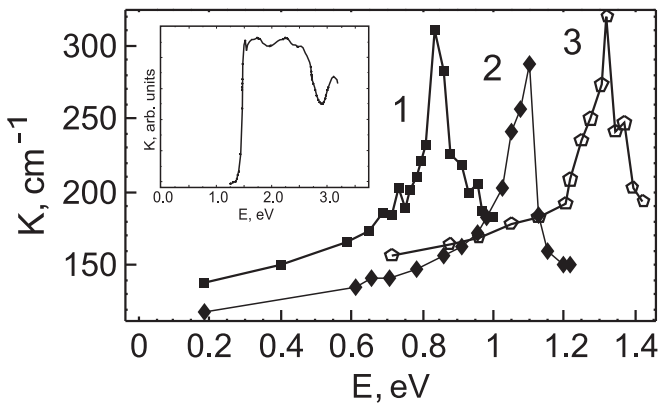}
\caption{The absorption spectra of CuO sample after the bobmardment with He$^+$ nuclei \cite{CuHe}.
Spectra 1 and 2 are obtained in plane polarizations $\mathbf{E}
\parallel [\overline{1}01]$ and $\mathbf{E} \perp [\overline{1}01]$, respectively. Spectrum 3 is obtained in unpolarized light. Solid lines are guides to the eye. Inset: the absorption spectra of CuO high quality single crystal.}
\end{figure}

In contrast to this natural picture, the examination of the spectra at the figure 1
leads us to the conclusion, that such a superposition principle for the light waves in
the system under consideration is violated. If we choose the polarizations $\mathbf{E}
\parallel [\overline{1}01]$ and $\mathbf{E} \perp [\overline{1}01]$ as the basic ones,
then the shape of the unpolarized light spectrum (curve 3 at figure 1) may only be
understood, if we assume, that the propagation properties of one of these basic waves
(or rather the corresponding system of two normal modes with elliptical polarization,
see Section 5) is modified in presence of the second wave, so as to give rise to the
peak at 1.3 eV and to hide the peaks at 0.9 eV and 1.1 eV.

In other words, the optical response of the system at a given frequency should depend
on the parameters of the incident light wave. This implies the optical nonlinearity of
the system. As a rule, the optical nonlinearity in non-magnetic media is described is
terms of the dielectric permittivity $\eps$, that depends on the intensity of the
incident light wave, $\eps = \eps(\w, |E|^2)$, where $\w$ is its frequency and $E$ --
the electric vector. However, in the present case we have to deal with a somewhat
subtler and less conventional type of nonlinearity, when the propagation of wave with
a given polarization depends on the intensity of {\it other} wave, propagating
simultaneously with this one.

We believe, that this unusual behaviour can be understood, assuming that the
bombardment of CuO sample with the fast He$^+$ nuclei leads to the formation of the
optical switching centers (OSC) -- the specific structures, that provide the optical
nonlinearity as discussed above. The basic ideas of the OSC model are discussed in the
two next sections.

\section{The model}

Consider a system, that contains the structures (hereafter called the centers) with
two equivalent metastable states $\ket{1}$, $\ket{2}$, separated by a potential
barrier, high enough to suppress the spontaneous transitions between them via
tunnelling and thermal activation. Thus, at low temperatures, the center is always
frozen in one of the two states. However, it may surmount the barrier and hop to the
other state when absorbing a portion of energy from an external energy reservoir.

The key points of the model are as follow: i) the center undergoes the transitions
$\ket{1}\leftrightarrows \ket{2}$ when absorbing a photon of an appropriate frequency
and ii) the transition $\ket{1}\rightarrow \ket{2}$ is allowed in a certain photon
polarization $p_1$ while the reverse transition $\ket{2} \rightarrow \ket{1}$ is
allowed in the polarization $p_2$, {\it orthogonal} to $p_1$. This is possible, if the
transitions involve an intermediate excited state $\ket{\psi^{\prime}}$ and the
transitions $\ket{i} \rightarrow \ket{\psi^{\prime}}$ are allowed in different
polarizations for $i = 1, 2$. Then the height of the barrier is $U = E_{\psi^{\prime}}
- E_{1,2}$, $E$ being the corresponding energy level (figure 2). Thus, the center acts
as an optical switch, which is triggered when irradiated by a light wave of an
appropriate polarization but does not feel the light of the complementary (orthogonal)
polarization.

Now, we examine the effects of interaction of the centers with light. Let us first
consider the case of the incident light wave with one of the special ($p_1$ or $p_2$)
polarizations, say $p_1$. Those centers, that are in the state $\ket{2}$ do not feel
this wave, while the centers in the state $\ket{1}$ absorb it and are excited to the
state $\ket{\psi^{\prime}}$. Then, these may either return to the state $\ket{1}$ or
switch to the state $\ket{2}$. The similar considerations hold for the light with the
polarization $p_2$. Given an arbitrary polarization, the wave function of the photons
may be written as
\begin{equation}\label{polar}
 \ket{p} = \cos \f \, \ket{p_1} + \sin \f \, \ket{p_2},
\end{equation}
where by assumption, two basic polarizations are orthogonal, $\bracket{p_1}{p_2} = 0$.
If the light is unpolarized, $\f$ is a random variable.
It follows from eq.(\ref{polar}), that
\begin{equation}\label{relI12}
  I_1/I_0 = \cos^2\f,\quad I_2/I_0 = \sin^2\f,
\end{equation}
where $I_0$, $I_{1,2}$ stand for the intensities of the incident wave and those of its
components with polarizations $p_{1,2}$, respectively.
Since the photons of both $p_1$- and $p_2$- polarization are present in this mixed
state in ratio $\cos^2 \f : \sin^2 \f$, the center will switch continuously
between the metastable states $\ket{1}$, $\ket{2}$: the photons with different polarizations throw the center from one potential well into the other.

Restricting ourself to the three-level model as discussed above, we may write down
the kinetic equations for the occupancy number of $\ket{1}$, $\ket{2}$,
$\ket{\psi^{\prime}}$ states, denoted $n_1$, $n_2$ and $n_3$, respectively:
\begin{align}
  \dot{n}_1 &= - W_{31} n_1 + W_{13} n_3, \label{kinetic1} \\
  \dot{n}_2 &= - W_{32} n_2 + W_{23} n_3, \\
  \dot{n}_3 &= W_{31} n_1 + W_{32} n_2 - (W_{13} + W_{23}) n_3,\\
  n_1 &+ \, n_2 + n_3 = 1, \label{kinetic4}
\end{align}
where $W_{ij}$ is the probability for a center to undergo the transition
$\ket{j}\rightarrow\ket{i}$. The equilibrium ($\dot{n}_{1,2} = 0$) relative numbers of
the centers in $\ket{1}$, $\ket{2}$ states are found to be:
\begin{align}
  n_1 &= \frac{W_{13} W_{32}}{W_{23} W_{31} + (W_{13} + W_{31}) W_{32}}\label{n1}, \\
  n_2 &= \frac{W_{23} W_{31}}{W_{23} W_{31} + (W_{13} + W_{31}) W_{32}}\label{n2}.
\end{align}
The excitation of the center from $i$-th potential well is stimulated with light of an
appropriate polarization. It follows from the symmetry arguments, that the transitions
$\ket{1,2} \rightarrow \ket{\psi^{\prime}}$ are governed by the same matrix element,
multiplied by the relative number of photons of that polarization (eq.\ref{relI12}),
whence:
\begin{align}
 W_{31} &= W \cos^2 \f, \label{WW1} \\
 W_{32} &= W \sin^2 \f.  \label{WW2}
\end{align}
The transition amplitude $W$ is found from the Fermi's golden rule:
\begin{equation}\label{W}
 W = \frac{4 \pi^2}{\hbar^2} \rho(\w_{12}) \left| \elem{\psi^{\prime}}{\mathbf{d}}{1,2} \right|^2,
\end{equation}
where $\mathbf{d}$ is the dipole moment operator and $\rho(\w_{12})$ is the spectral
density of radiation at the frequency of the transition, the latter being proportional
to the height of the potential barrier, separating the metastable states of the center
(figure 2):
\begin{equation}\label{barrier}
\hbar\,\w_{12} \,\sim \,U.
\end{equation}
As for the transitions $\ket{\psi^{\prime}}\rightarrow \ket{1,2}$,
these may be both spontaneous and stimulated, and their total probabilities
$W_{13}$, $W_{23}$ include the contributions of both channels. Thus, we write:
\begin{equation}\label{WW3}
W_{13} = W_{13}^{\prime} + W_{13}^{\prime\prime} \cos^2 \f,
\end{equation}
and similarly for $W_{23}$, where the probability $W_{13}^{\prime}$ of the spontaneous
transition obviously does not depend on the polarization of incident photons and
$W_{13}^{\prime\prime}$ corresponds to the transition, induced with photons of $p_1$
polarization, their relative number being $\cos^2 \f$.

Having in mind to account for the polarization anomalies in experimental spectra
(figure 1), as discussed in previous section, we are mainly interested in the general
properties of the polarization dependence of the stationary occupancy numbers
$n_{1,2}(\f)$. Taking into account eqs.(\ref{WW1}-\ref{WW3}) and assuming
$W_{i3}^{\prime} = W^{\prime}$,
$W_{i3}^{\prime\prime} = W^{\prime\prime}$ for $i = 1, 2$
to make the general formulae (\ref{n1},\ref{n2}) more tractable, one gets:
\begin{subequations}\label{n12}
\begin{align}
 n_1 = \widetilde{n}(\f) + \frac{\sin^2 \f}{1+ (\widetilde{W}/W^{\prime}) \cos^2 \f \sin^2\f}, \\
n_2 = \widetilde{n}(\f) + \frac{\cos^2 \f}{1+ (\widetilde{W}/W^{\prime}) \cos^2 \f \sin^2\f},
\end{align}
\end{subequations}
where $\widetilde{W} = W + 2 W^{\prime\prime}$ and
\begin{equation}\label{n0}
\widetilde{n}(\f) = \frac{(W^{\prime\prime}/W^{\prime})\,\cos^2\f \sin^2\f}{1+ (1 + 2 W^{\prime\prime}/W^{\prime}) \cos^2\f \sin^2\f}.
\end{equation}
\begin{figure}[t]\label{pjt}
\includegraphics[width=\linewidth]{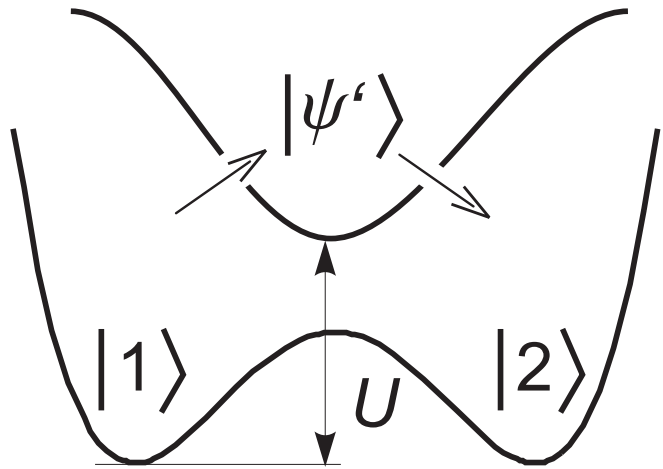}
\caption{Schematic picture of the optical switching transition.  The center is excited
from its ground state $\ket{1}$ to the intermediate state $\ket{\psi^{\prime}}$,
absorbing the photon with polarization $p_1$, and then may either relax back to its
initial state or switch to another ground state $\ket{2}$.}
\end{figure}
The function $\widetilde{n}(\f)$ is symmetric in the number of the photons with $p_1$,
$p_2$ polarization, while the second terms in rhs. of eqs.(\ref{n12}) show, that the
equilibrium number $n_1$ of the centers, sensitive to the photons with $p_1$
polarization depends on the fraction of the photons with $p_2$ polarization in the
beam (\ref{relI12}), and vice versa. 

Without the latter assumption of equal
probabilities $W_{i3}^{\prime}$'s, $W_{i3}^{\prime\prime}$'s, more complicated
expressions are obtained instead of (\ref{n12}), but the qualitative results still
remain the same: $n_1 = 0$, $n_2 = 1$ under the irradiation by light with $p_1$
polarization ($\f = 0$) and $n_1 = 1$, $n_2 = 0$ for the $p_2$ polarization ($\f =
\pi/2$). The reason is that due to the switching described above, the center acquires
a finite (albeit low) probability of {\it irreversible} escape from one state to the
other under the irradiation with $p_1$ or $p_2$ polarization and thus all centers are
switched for a long enough time. For other polarizations, both $n_1$ and $n_2$ are
non-zero, because once switched, the center has a finite probability to switch back
due to the presence of the photons with an appropriate polarization in the incident
wave. 

\section{The optical response}

The switching transitions give rise to the resonant optical absorption at the
frequency $\w_{12}$, eq.(\ref{barrier}). We next put forward the general arguments to
propose a plausible form of the dielectric tensor with regard to the switching
transitions. To be specific, we consider the transmission of light through the film,
containing the switching centers. In the case of normal incidence, the electric vector
of the light wave lies in plane of the film. For the sake of simplicity, it will be
assumed, that the in-plane part of the dielectric permittivity tensor $\hat{\eps}$ is
diagonal within the basis of the polarization states $\ket{p_{1,2}}$:
\begin{align}\label{epsilon}
 \hat{\eps} &= \left( {\begin{array}{*{20}c}
                      {\eps_1} & {0} \\
                         {0}   & {\eps_2}\\
                      \end{array}} \right),
\end{align}
i.e. the corresponding plane waves are the normal modes for the system. In general,
$\eps_{1,2}$ contain the contributions from different optical transitions. E.g., as
far as the optical transitions between the discrete levels are considered, it is
common to express the dielectric permittivity matrix elements as a sum of lorentzian
functions:
\begin{equation}\label{lor}
 \mathcal{L}\left(\w;\, \{\w_i, I_i, \gamma_i\}\right)\,=\, 1 \,+\, \frac{1}{4 \pi} \sum\limits_i
 \frac{I_i}{\w^2 - \w_i^2 + i \gamma\, \w_i},
\end{equation}
where $\w_i$, $I_i$ and $\gamma_i$ are the resonance frequency, intensity and damping
factor of $i-$th optical transition, respectively. However, we emphasize, that the
specific form of the spectral line is not important in what follows and the Lorentzian
function is chosen rather arbitrarily.

Now, we focus on those contributions to the $\hat{\eps}$ tensor,
that are related to the switching transitions $\ket{1} \leftrightarrows \ket{2}$ . By
definition, their intensities are:
\begin{subequations}\label{I12}
\begin{align}
  I_{1 \rightarrow 2} = \hbar \w_{12} P_1, \\
  I_{2 \rightarrow 1} = \hbar \w_{12} P_2,
\end{align}
\end{subequations}
where $P_i$ is the probability for the photon with polarization $p_i$ to be absorbed
by a center:
\begin{subequations}\label{P12}
\begin{align}
  P_1 &= W_{21} n_1, \\
  P_2 &= W_{12} n_2.
\end{align}
\end{subequations}
Here $W_{ij}$ is the amplitude of the transition $\ket{j}\rightarrow\ket{i}$, $W_{ij}
\propto W_{i\,3} W_{3\,j}$ (eqs.\ref{kinetic1}-\ref{kinetic4}) and $n_{1,2}$ are given
by eqs.(\ref{n12}). Taking into account (\ref{P12}), the substitution of (\ref{I12})
into (\ref{lor}) makes it possible to write down the dielectric permittivity tensor
(\ref{epsilon}) in the form:
\begin{align}\label{epstens}
 \hat{\eps} &=
 \left( \begin{array}{*{20}c}
 {\eps_1 + \chi \cdot n_1(\f)}  & {0} \\
      {0} & {\eps_2 + \chi \cdot n_2(\f)} \\
\end{array} \right),
\end{align}
where $\f$ is the polarization angle of incident light wave relative to the basis
$\ket{p_{1,2}}$ (\ref{polar}). All contributions to the conventional linear optical
response of the system are absorbed in $\eps_{1,2}$ terms, whereas the angle-dependent
terms describe the contribution of the switching transitions. The angular dependence
of $\hat{\eps}$ matrix elements arises due to the feedback between the extinction of a
certain normal mode and the intensity of other normal mode with complementary
polarization. Such a feedback, provided by OSC's, is described by eqs.(\ref{n12},\ref{I12},\ref{P12}).


The dielectric permittivity tensor $\hat{\eps}$, eq.(\ref{epstens}), describes the optical
response of a nonlinear system, which parameters vary with the polarization of
incident light.
Taking into account eqs.(\ref{relI12},\ref{n12},\ref{epstens}), we may write:
\begin{subequations}\label{ourNonlin}
\begin{align}
 \hat{\eps}_{11} = \hat{\eps}_{11}(\w, I_2/I_0),\\
 \hat{\eps}_{22} = \hat{\eps}_{22}(\w, I_1/I_0).
\end{align}
\end{subequations}
This type of nonlinearity differs from the conventional one, 
\begin{equation}\label{Nonlin}
 \hat{\eps} = \hat{\eps}(\w, I_0), 
\end{equation}
when the nonlinear effects are governed by the intensity $I_0$ of incident light.

All terms in eq.(\ref{epstens}) are functions of frequency, which explicit form should be
specified proceeding from a relevant microscopic theory. E.g., the function $\chi$
will depend on the profile of the potential barrier (figure 2) via the transition
amplitudes $W$, $W^{\prime}$, $W^{\prime\prime}$  (\ref{n12}), etc. It is not our
intention here to develop such a detailed theory, as it would require a number of
additional model assumptions. Instead, we stress, that crucial is the angular
dependence in eq.(\ref{epstens}) itself, and not a special form of the functions
$\eps_{1,2}(\w)$ and $\chi(\w)$. However, taking into account, that $\chi(\w)$
describes the hopping of the OSC over the potential barrier, it seems reasonable to
assume, that it should exhibit a resonance at the frequency $\w_{12}$, eq.(\ref{barrier}).

Given the dielectric permittivity tensor, it is straightforward to derive the
absorption coefficient $K$ of the film with thickness $d$ for an arbitrary
polarization of light (for simplicity, the effects of boundary reflections are
neglected):
\begin{equation}\label{bouguer}
 I_{out}/I_{in} = \exp(- K d) = U_1\,\cos^2 \f\,+\,U_2\,\sin^2 \f,
\end{equation}
where $I_{in}$ and $I_{out}$ are the intensities of incident and transmitted wave,
respectively,  $U_i = \exp\left(- 2 (\w/c) n_i^{\prime\prime} d\right)$ and $n_i =
n_i^{\prime} + i n_i^{\prime\prime} = \eps_i^{1/2}$ are refractive indices, $\eps_i$'s
being the eigenvalues of the dielectric permittivity tensor. The absorption
coefficient for unpolarized light is obtained by averaging over the polarizations:
\begin{equation}\label{unpol}
 K_{unpol.} = \frac{1}{2\,\pi} \int\limits_0^{2\,\pi} K(\f) \, d\f.
\end{equation}

The key result of the present model is that the contribution of the $\chi(\w)$ term in the dielectric tensor, eq.(\ref{epstens}), to the
absorption coefficient is effective only in presence of both normal modes with
polarizations $p_1$ and $p_2$. This may be seen, noting, that the absorption
coefficient and other functions of optical response are expressed as the functions of
the ''matrix elements'': 
\begin{equation}
 \elem{\mathbf{E}}{\hat{\eps}}{\mathbf{E}} = |E_0|^2\,\left((\eps_1 + \chi\,n_1(\f))\,\cos^2 \f + (\eps_2 + \chi\,n_2(\f))\,\sin^2 \f \right),
\end{equation}
where
$\mathbf{E} = E_0 (\cos \f, \sin \f)$ is the electric vector of incident light.
With $n_{1,2}$ given by eqs.(\ref{n12}), it is obvious, that the contribution of $\chi(\w)$ to this quantity vanishes for $\f$ close to $0$ and $\pi/2$.

\begin{figure}[t]
\includegraphics[width=\linewidth,angle=0]{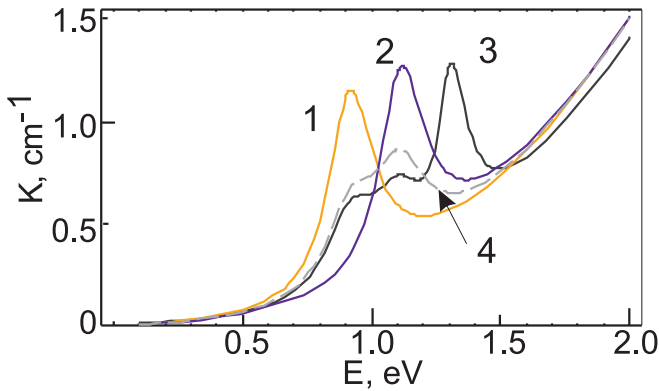}
\caption{The absorption coefficient, obtained with eqs.(\ref{bouguer},\ref{unpol}). Curves
1 and 2 depict the response in two plane polarizations, $\f = 0$ and $\f = \pi/2$,
respectively. The curve 3 simulates the spectrum, obtained in unpolarized light.
Noteworthy is the novel peak at 1.3 eV, not observed in two previous polarizations.
The dashed curve 4 is the unpolarized light spectrum of a conventional medium and is
obtained using the trivial counterpart of the dielectric tensor, eq.(\ref{epstens}), where the angle-dependent terms are omitted.}
\end{figure}

For illustrative purposes, we consider a simple case, when $\eps_{1,2}(\w)$ and
$\chi(\w)$ show well resolved resonances at different frequencies. All contributions
in the $\hat{\eps}$ tensor (\ref{epstens}) have been simulated with the Lorentzian
functions (\ref{lor}). The ''conventional'' contributions due to some optical
transitions in the medium are chosen to be $\eps_{1,2}(\w) = \mathcal{L}\left(\w;\,
\{\w_{1,2}, 1.1, 0.2\}, \{2.5, 20.2, 2.0 \}\right)$ with $\w_1 = 0.9$ eV and $\w_2 =
1.1$ eV. The OSC contribution have been taken in the form: $\chi(\w) =
\mathcal{L}\left(\w;\,\{\w_3, 5.5, 0.1\}\right)$ with a sharp resonance at $\w_3 =
1.3$ eV. The ratios $\widetilde{W}/W^{\prime}$ and $W^{\prime\prime}/W^{\prime}$ in
eq.(\ref{n12}) are found to show a minor effect on the final results when varying in a
broad range and are (arbitrarily) set equal to $1.0$ and $0.2$, respectively. As shown
in figure 3, the absorption spectra are different in two basic polarizations (curves 1
and 2), that is typical of the anisotropic media. The sharp peaks, observed in two
polarization, are also visible as the slight features at the curve 3, simulating the
unpolarized light spectrum. In intermediate polarizations, the switching transitions
come into play, and the curve 3, simulating the natural light spectrum, displays a
novel strong resonance at 1.3 eV, that emerges  due to $\chi(\w)$ and is absent in
both plane polarizations.
\begin{figure}[t]
\includegraphics[width=0.6\linewidth,angle=0]{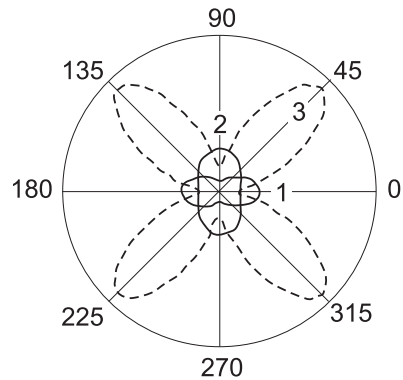}
\caption{The absorption coefficient $K(\w, \f)$, eq.(\ref{bouguer}), at selected
frequencies as a function of polarization angle $\f$. 1: $\w = \w_1$. 2: $\w =
\w_2$. 3: $\w = \w_3$ (see text).}
\end{figure}
This behaviour is clearly at variance to what is expected for any conventional system,
where the unpolarized spectrum should be merely a superposition of the spectra,
observed at two different polarizations. For comparison, such a ''usual'' unpolarized
spectrum, obtained from eqs.(\ref{epstens},\ref{unpol}) with $\chi \equiv 0$ is also shown
at figure 3 (curve 4). Of course, the numerical quantities in this example do not
pretend to be the fitting parameters, relevant to the experimental spectra at figure
1.

The polarization dependence of the model spectra is shown at figure 4. It is seen,
that the indicatrix of the absorption coefficient at the resonance frequency $\w_1$
behaves as $K(\w_1,\f) \sim \cos^2 \f$ and is maximum at the polarization angle $\f =
0$, while at the resonance frequency $\w_2$, $K(\w_2,\f) \sim \sin^2 \f$ is maximum at
$\f = \pi/2$. The absorption coefficient at the switching frequency $\w_3$ is minimum
in both of these basic polarizations, but shows a pronounced maximum at $\f = \pi/4$.
The reason is that the switching of OSC requires both normal modes to be excited, and
this condition is optimally fulfilled in $\pi/4$- polarization.

\section{Discussion}

The theoretical interpretation of the optical data on cupric oxide CuO is particularly
difficult because of its low symmetry monoclinic tenorite structure. For the geometry
of experiment, as described in Section 2, the tensor $\hat{\eps}$ of an ideal single
crystalline sample is decomposed into the scalar $\eps_b$ and the in-plane part
\cite{kuzmenko}:
\begin{align}\label{ac}
 \hat{\eps}_{ac} &=
 \left( \begin{array}{*{20}c}
 {\eps_{xx}} & {\eps_{xz}}\\
 {\eps_{zx}} & {\eps_{zz}}\\
\end{array} \right),
\end{align}
where the orthogonal coordinate axes $\mathbf{x}$, $\mathbf{z}$ lie in plane of the
film (the crystallographic $ac$ plane). The normal modes are the transverse waves with
elliptic polarizations, propagating with (complex) refractive indices $n_{1,2}$:
\begin{equation}
2 n^2_{1,2} = (\eps_{xx} + \eps_{zz}) \pm ((\eps_{xx} - \eps_{zz})^2 + 4 \eps_{xz}\,\eps_{zx})^{1/2}.
\end{equation}
Within the disordered sample, the propagation direction of light may deviate from that
of the $\mathbf{b}$ axis because of the scattering on the nonuniformities, induced
with the fast particle bombardment. The excitation of both transverse and longitudinal
modes can therefore be expected. Thus, more realistic theoretical approaches should
deal with the non-diagonal tensors of a general form, as distinct from the model
assumption, used in eq.(\ref{epstens}), that have been made for the illustrative
purposes.

However, we would like to emphasize, that the interpretation of the experimental
spectra at figure 1 does not reduces to the accurate analysis of the normal modes of
some disordered low-symmetry linear medium. In fact, two different problems should be
distinguished here: those of the explanation of $K(E)$ dependence for a certain
polarization, and those related to the relative positions of the three spectra. While
the form of the $\hat{\eps}(\w)$ tensor determines the shape of the optical spectrum
for any particular polarization, the problem of the abnormal polarization dependence
of the spectra is beyond the scope of a linear response theory, as have been discussed
in Section 2. We believe, that it can be resolved only assuming the explicit
dependence of the dielectric tensor on the polarization of incident light.

The polarization-dependent switching effect, proposed in previous sections, provides
the required general form of the $\hat{\eps}$ tensor, eq.(\ref{epstens}), and makes it
possible to realize, how in principle the abnormal polarization dependence of the
absorption spectra of CuO (figure 1) could be understood. As the theoretical spectra
at figure 3 have been obtained with rather arbitrarily chosen numerical values of the
model parameters, our present results are preliminary and allow but a qualitative
comparison with experiment. However, the polarization dependence of the calculated
spectra (figure 4) stems from general formulae (\ref{n12},\ref{epstens}) and is not related to
some particular choice of the fitting parameters. Overall, the spectra at the figure 3 capture the most essential features of experiment, except that the curve 3, simulating the natural light spectrum, displays the traces of the peaks 1 and 2, in contrast to the corresponding spectrum 3 at figure 1. We believe, that rather large experimental errors, noted in \cite{CuHe}, make it
difficult to decide, whether some slight features can in fact be resolved at the broad
tails of the experimental spectra.

The natural question arises about the microscopic physics of OSC. To our opinion,
under certain conditions the Jahn-Teller clusters may be regarded as their possible
prototypes. To be specific, we consider $E - b_1 - b_2$ - problem \cite{bersuker},
that arises e.g. in the Jahn-Teller theory of the square cluster with the $D_{4h}$
point symmetry when its electron ground state orbital $E_u\{x, y\}$ doublet is coupled
to $b_{1g}$ and $b_{2g}$ local lattice modes. The vibronic interaction causes the
spontaneous symmetry breaking of the cluster and the profile of its adiabatic
potential along the active Jahn-Teller coordinate is shown schematically at figure 2.
E.g., if the vibronic coupling with $b_{1g}$ mode dominates, the cluster acquires one
of two equivalent static $b_{1g}$ (rhombic) distortions (the ground state average
value of the corresponding normal coordinate being non-zero: $\langle Q_{b1g} \rangle
= \pm Q_{b1g}^0$). The resulting vibronic ground state wave functions of the cluster
frozen in one of the two potential wells have the symmetry properties: $\Psi_{\pm}
\sim x, y$. If there is also some excited high-symmetry state $\psi^{\prime} \sim
A_{1g}$, then the photon-activated switching between the two Jahn-Teller distorted
configurations, involving the intermediate $\psi^{\prime}$ state, can be considered
(figure 2). Noteworthy is that the dipole transition matrix elements
$\elem{\psi^{\prime}}{(\mathbf{E}\cdot\mathbf{d})}{\Psi_{\pm}}$ are allowed in
orthogonal polarizations, $\mathbf{E}\parallel\,x$ and $\mathbf{E}\parallel\,y$,
respectively. Thus, the Jahn-Teller cluster may have all essential ingredients of the
OSC as discussed in Section 3. In principle, the similar arguments may also be
relevant for other cases of a spontaneous symmetry breaking and the role of the
lattice subsystem is not crucial. In this connexion we note an interesting idea, put
forward in \cite{nonrigid}, about the possibility of the ''purely electronic''
Jahn-Teller effect in strongly correlated systems, that consists in the correlational
shift of the electron shells, similar to the conventional vibronic
distortion of the lattice.

The Jahn-Teller effect may in fact play an important role in the physics of cuprates.
In particular, it was shown in a series of papers \cite{mosk,physB,physchem,pss}, that
many peculiarities of the optical and structural properties of CuO and other cuprates
allow a consistent interpretation within the model of polar electron-hole
pseudo-Jahn-Teller (PJT) centers, which underlying physics is dominated by the
near-degeneracy of $A_{1g}$ and $E_u$ molecular orbitals. The authors note, that the
favourable conditions for the nucleation of PJT centers is  the chemical doping, fast
particles bombardment or other perturbations, that may locally disturb the stability
of the parent system. Thus, we believe the PJT centers in cuprates to be a plausible
candidates to the OSC's, considered in this work.

\section{Summary}

In conclusion, the aim of the present work is to explain the polarization anomalies
observed in the optical absorption spectra of cupric monoxide CuO after the
bombardment with He$^+$ nuclei \cite{CuHe}: the spectra, obtained in unpolarized light
were found to display new features, that are not observed in two different plane
polarizations (figure 1). Note, that for such an effect to be revealed, the
comparative analysis of at least three spectra in different polarizations is required.
To the best of our knowledge, no similar results have been ever observed in experiment or investigated theoretically.
Leaving aside a number of involved problems, concerning the electronic structure of
cupric monoxide and its changes under the bombardment, we addressed the question: how,
at least qualitatively, these polarization anomalies could be explained. To this end
we have proposed the model of a nonlinear system, which dielectric permittivity $\eps
= \eps(\w, \f)$, eqs.(\ref{epstens},\ref{ourNonlin}), depends explicitly on the polarization angle $\f$ of light,
propagating through the system. This type of nonlinearity differs from the
conventional one, $\eps = \eps(\w, |E|^2)$, where $|E|^2$ is the intensity of light.
In our model, such a nonlinearity arises due to the special structures, the optical
switching centers (OSC), which behaviour is as follows: i) The center has two
metastable states, $\ket{1}$, $\ket{2}$; ii) it can hop from $\ket{1}$ to $\ket{2}$,
absorbing the photons with polarization $p_1$ and iii) back from $\ket{2}$ to
$\ket{1}$, absorbing the photons with polarization $p_2$, {\it orthogonal} to $p_1$.
These switching transitions $\ket{1}\leftrightarrows\ket{2}$ give rise to the novel
channel of optical absorption, that is only effective in presence of waves with both
polarizations. This explains, how the new spectral features, not observed in two orthogonal polarizations $p_1$ and $p_2$, can emerge in unpolarized light.

We propose, that the systems with the properties, typical of OSC, can be found among
the Jahn-Teller clusters. Taking into account, that the polar Jahn-Teller centers have
been shown to play an important role in the optics of copper oxides
\cite{CuHe,CuEl,CuAzot,mosk,physB,physchem,pss}, we believe, that the present model
may indeed capture some interesting and yet little explored physics of cuprates. More
detailed theory, that allows the quantitative comparison with experiment, will be
deferred for other publication.

\begin{acknowledgments}
The author is indebted to Drs. Yu.P. Sukhorukov and Yu.D. Panov for many
valuable comments. The discussions with Profs. N.G. Bebenin and V.Y. Shur are also
acknowledged. The work is supported by Grant 04-02-96068 RFBR URAL 2004.
\end{acknowledgments}

\end{document}